\documentclass[useAMS,usenatbib]{mn2e}

\usepackage{lastpage}
\usepackage{times}
\usepackage{amsmath,amsfonts,graphicx}
\usepackage{color}

\newcommand{\mj}{\mathrm{M}_\mathrm{J}}
\newcommand{\msun}{\mathrm{M}_\odot}
\newcommand{\yr}{\mathrm{yr}}

\renewcommand{\vec}[1]{\boldsymbol{#1}}

\title[The planets around NN~Ser: still there]{The planets around
NN~Serpentis: still there%
\thanks{Partly based on observations collected at the European 
Southern Observatory, La Silla and Paranal, Chile (programmes
087.D-0593, 090.D-0277 and 091.D-0444)}
}
\author[Marsh, Parsons, Bours, Littlefair, Copperwheat,
Dhillon, Breedt, Caceres \& Schreiber]{T.\ R.\ Marsh$^{1}$\thanks{E-mail:
t.r.marsh@warwick.ac.uk}, S.\ G.\ Parsons$^{2}$, M.\ C.\ P.\ Bours$^{1}$,
  S.\ P.\ Littlefair$^{3}$, C.\ M.\ Copperwheat$^{4}$
\newauthor
V.\ S.\ Dhillon$^{3}$, E.\ Breedt$^1$, C.\ Caceres$^{2}$ and M.\ R.\ Schreiber$^{2}$.\\
$^{1}$Department of Physics, University of Warwick,
Gibbet Hill Road, Coventry, CV4 7AL, UK\\
$^{2}$Departamento de Fisica y Astronomia, Faculdad de Ciencias,
Universidad de Valparaiso, Chile\\
$^{3}$Department of Physics and Astronomy, University of Sheffield, 
Sheffield, S3 7RH, UK\\
$^{4}$Astrophysics Research Institute, Liverpool John Moores University, Twelve Quays House, Birkenhead, CH41 1LD, UK
}
\begin{document}

\date{Accepted ----. Received ----; in original form ----}

\pagerange{\pageref{firstpage}--\pageref{lastpage}} \pubyear{2012}

\maketitle

\label{firstpage}

\begin{abstract}
  We present 25 new eclipse times of the white dwarf binary NN~Ser
  taken with the high-speed camera ULTRACAM on the WHT and NTT, the RISE
  camera on the Liverpool Telescope, and HAWK-I on the VLT to test the
  two-planet model proposed to explain variations in its eclipse times
  measured over the last 25 years. The planetary model survives the test with
  flying colours, correctly predicting a progressive lag in eclipse times of
  36 seconds that has set in since 2010 compared to the previous 8
  years of precise times. Allowing both orbits to be eccentric, we find
  orbital periods of $7.9 \pm 0.5\,$yr and $15.3 \pm
    0.3\,$yr, and masses of $2.3\pm 0.5\,\mj$ and $7.3\pm
    0.3\,\mj$. We also find dynamically long-lived orbits consistent with the
  data, associated with 2:1 and 5:2 period ratios. The data scatter by
  $0.07$ seconds relative to the best-fit model, by some margin the
  most precise of any of the proposed eclipsing compact object planet
  hosts. Despite the high precision, degeneracy in the orbit fits prevents a
  significant measurement of a period change of the binary and of $N$-body
  effects. Finally, we point out a major flaw with a previous dynamical
  stability analysis of NN~Ser, and by extension, with a number of analyses of
  similar systems.
\end{abstract}

\begin{keywords}
(stars:) binaries (including multiple:) close -- (stars:) binaries: eclipsing
  -- (stars:) white dwarfs -- (stars:) planetary systems 
\end{keywords}

\section{Introduction}
The discovery of hundreds of planets around stars other than the Sun has
alerted researchers to the possible influence of planets in a wide variety of
circumstances. Amongst these are the spectacular Kepler discoveries of planets
transiting across both stars of the tighter binary systems around which they
orbit \citep{Doyle2011Sci333.1602,Welsh2012Nat481.475}. The transits in these
systems leave no doubt as to the existence of planets in so-called ``P-type''
orbits \citep{Dvorak1986AA167.379} around binaries. Even before the Kepler
discoveries there was evidence for planets around binaries from timing
observations of a variety of systems where the presence of planets is
indicated through light travel time (LTT) induced variations in the times of
eclipses. This method has led to claims of planetary and/or sub-stellar
companions around hot subdwarf/M dwarf binaries
\citep{Lee2009AJ137.3181,Qian2009ApJ695L163}, white dwarf/M dwarf binaries
\citep{Qian2009ApJ706L96,Qian2010MNRAS401L34,Beuermann2010AA521L60}, and
cataclysmic variables
\citep{Beuermann2011AA526.53,Qian2011MNRAS414L16,Potter2011MNRAS416.2202}. In
all the cases cited one of the binary components is evolved which helps
observationally because the evolved star is hot and relatively small, leading
to sharply-defined, deep edges in eclipse light curves which make for precise times.

Planets discovered through timing complement those found in radial velocity
and transit surveys as they are easier to discover the larger (and thus longer
period) their orbits are. The existence of planets around evolved stars raises
interesting questions as to whether the planets are primordial and managed to
survive the evolution of the host binary, or whether they instead formed from
material ejected during the course of stellar evolution
\citep{Beuermann2011AA526.53,Veras2012MNRAS422.1648,Mustill:2013arXiv1309.3881M},
and may also place unusual constraints upon the binary's evolution
\citep{PortegiesZwart2013MNRAS429.45}.

The Kepler discoveries prove that circumbinary planets exist, but when it
comes to those discovered through timing, the reality of the planets is not
clear-cut. The history of the field is not encouraging in this respect. For
instance, the orbits measured for the white dwarf/M dwarf binaries NN~Ser and
QS~Vir by \citet{Qian2009ApJ706L96} and \citet{Qian2010MNRAS401L34} were both
ruled out as soon as new data were acquired \citep{Parsons2010MNRAS407.2362},
as were the two-planet orbits proposed by \cite{Lee2009AJ137.3181} for the
sdB+dM binary HW~Vir \citep{Beuermann2012AA543.138}. Likewise, some multiple
planet systems claimed from timing studies \citep{Qian2011MNRAS414L16} have
had problems with long-term dynamical stability
\citep{Horner2011MNRAS416L11,Hinse2012MNRAS420.3609,Potter2011MNRAS416.2202}.
These are serious issues because there is no independent evidence yet for the
existence of the various third-bodies suggested by timing, while the mere fact
that timing variations can be fitted by planetary models is not entirely
persuasive, since with enough extra bodies the process is akin to fitting a
Fourier series, and any set of data can be matched. At present, the main rival
model for the period changes is one in which they are caused by fluctuations
in the gravitational quadrupolar moments of one or both stars
\citep{Applegate1992ApJ385.621}. In some cases this appears to fail on
energetic grounds \citep{Brinkworth2006MNRAS365.287}, and at the moment this
constitutes the only, rather indirect, independent support for the planetary
hypothesis for the eclipse timing variations of compact binary stars, although
artefacts of measurement, such as wavelength-dependent eclipse
  timings, are a possible issue in the case of accreting systems
\citep{Gozdziewski2012MNRAS425.930}.

Useful scientific hypotheses have predictive power. So far the planetary
explanation of LTT variations has fared poorly on this basis. In this paper we
present new observations of the system NN~Ser which is currently the most
convincing example of an LTT-discovered planetary system around a close binary
star. Our aim is to see whether the planetary model developed by
\citet{Beuermann2010AA521L60} can withstand the test of new data. NN~Ser is a
white dwarf/M dwarf binary with an orbital period $P = 3.1\,$hours which was
discovered to eclipse by \citet{Haefner1989AA213.15}. The combination of a hot
white dwarf and low mass M dwarf ($0.111\,\msun$,
\citealt{Parsons2010MNRAS402.2591}), allows the white dwarf to dominate its
optical flux completely, giving very deep, sharply-defined eclipses which
yield precise times. The very low mass of the M dwarf is an important feature
since its low luminosity greatly restricts the effectiveness of
\citet{Applegate1992ApJ385.621}'s period change mechanism, as pointed out by
\citet{Brinkworth2006MNRAS365.287}, who first detected period changes in
NN~Ser.  \citeauthor{Brinkworth2006MNRAS365.287} interpreted the period
changes as a sign of angular momentum loss, but \citet{Beuermann2010AA521L60}
reanalysed an early observation of NN~Ser from the VLT and were able to show
that the orbital period was not simply changing in one direction but had shown
episodes of lengthening as well as shortening. They showed that the timing
variations could be well explained if there were two objects of minimum mass
$6.91\,\mj$ and $2.28\,\mj$ in orbit around the binary. This nicely solved the
problem that the period changes appeared to be much larger than expected on
the basis of the angular momentum mechanisms thought to drive binary evolution
\citep{Brinkworth2006MNRAS365.287,Parsons2010MNRAS402.2591}.

Of all the planets discovered through timing around binaries, those around
NN~Ser are arguably the most compelling because the data quality is so high
with the best times having uncertainties $< 0.1\,\sec$, because it is a
well-detached binary with an extremely dim main-sequence component, and since
the two planet model fits the eclipse times almost perfectly
\citep{Beuermann2010AA521L60}.  NN~Ser thus provides us with a chance to see
if the planet model is capable of predicting eclipse arrival times in
detail. This was the motivation behind this study.

Shortly after submitting this paper, another paper presenting eclipse times of
NN~Ser appeared \citep{Beuermann2013AA555A.133B}. We have elected not to
update our paper with their times to render a comparison with their results
more independent. We have included such a comparison in section~\ref{sec:new}.

\section{Observations and their Reduction}

We observed 25 eclipses of NN~Ser, over the period 25
February 2011 to 26 July 2013, extending the
baseline of the times presented in \citet{Beuermann2010AA521L60} by $3\,$years (Table~\ref{tab:obslog}).
\begin{table*}
\centering
\begin{minipage}{140mm}
\caption{New eclipse times of NN~Ser \label{tab:obslog}}
\begin{tabular}{ccccll}
  \hline
Cycle &    BMJD(TDB)    & Error ($1\,\sigma$)  &  Sampling & Tel/Inst   & Comments \\
      &      (days)    & (seconds)           & (seconds) &            &
      Transparency, seeing, etc.\\
\hline
61219 & 55307.4003018  & 0.084  & 3.0 & NTT/UCAM       & 
Update of time listed in \protect\citet{Beuermann2010AA521L60}.\\
61579 & 55354.2291437  & 0.064  & 2.6 & NTT/UCAM       & 
Update of time listed in \protect\citet{Beuermann2010AA521L60}.\\ 
63601 & 55617.2511773  & 0.341  & 6.0 & LT/RISE    & Clear, seeing $1.8$''.\\ 
63816 & 55645.2184078  & 0.500  & 6.0 & LT/RISE    & Clear, $2$''.\\ 
64032 & 55673.3157097  & 0.132  & 3.0 & NTT/UCAM   & Clear, $1.5$'', bright
Moon; $u'$, $g'$, $r'$.\\ 
64054 & 55676.1774753  & 0.402  & 6.0 & LT/RISE    & Clear, $2$''.\\ 
64322 & 55711.0389457  & 0.397  & 6.0 & LT/RISE    & Clear, $2$''.\\ 
64330 & 55712.0795926  & 0.057  & 2.3 & NTT/UCAM   & Clear,
$1.5$''; $u'$, $g'$, $r'$.\\ 
64575 & 55743.9492287  & 0.369  & 6.0 & LT/RISE    & Clear, $2$''.\\ 
64836 & 55777.9001514  & 0.347  & 5.0 & LT/RISE    & Clear, $2$''.\\ 
65992 & 55928.2728113  & 1.134  & 5.0 & LT/RISE    & Variable, $3$''.\\ 
66069 & 55938.2889870  & 0.256  & 3.4 & WHT/UCAM   & Cloudy, $1$'', bright
Moon, twilight; $u'$, $g'$, $r'$.\\ 
66092 & 55941.2808293  & 0.062  & 2.0 & WHT/UCAM   & Clear,
$1.5$''; $u'$, $g'$, $r'$.\\ 
66545 & 56000.2071543  & 0.425  & 5.0 & LT/RISE    & Clear, $\sim 1.8$''.\\ 
66868 & 56042.2230409  & 0.035  & 2.0 & WHT/UCAM   & Clear,
$2$''; $u'$, $g'$, $i'$.\\ 
66905 & 56047.0360108  & 0.080  & 2.0 & WHT/UCAM   & Clouds on
ingress and egress,
$2$''. Caution! See text.\\ 
67581 & 56134.9702132  & 0.421  & 5.0 & LT/RISE   & Clear, $2$'' \\ 
67903 & 56176.8560256  & 0.034  & 2.0 & WHT/UCAM   & Clear, $1$'',
twilight; $u'$, $g'$, $r'$. \\ 
67934 & 56180.8885102  & 0.044  & 2.1 & WHT/UCAM   & Clear,
$2$''; $u'$, $g'$, $r'$.\\ 
69067 & 56328.2693666  & 0.536  & 5.0 & LT/RISE    & Clear, $2.5$'' \\ 
69291 & 56357.4073373  & 0.657  & 7.0 & VLT/HAWK-I & Clear, 1'', twilight.\\ 
69298 & 56358.3178846  & 0.245  & 7.0 & VLT/HAWK-I & Clear, $0.5$''.\\ 
69336 & 56363.2609298  & 0.506  & 5.0 & LT/RISE    & Cloudy, $1.8$'' \\ 
69597 & 56397.2118717  & 0.491  & 7.0 & VLT/HAWK-I & Clear, $1$''. \\ 
69598 & 56397.3419520  & 0.392  & 7.0 & VLT/HAWK-I & Clear, $0.8$''.\\ 
70287 & 56486.9672059  & 0.037  & 2.4 & WHT/UCAM & Clear, $0.9$'; $u'$, $g'$, $i'$'.\\ 
70387 & 56499.9752252 & 0.041  & 2.1  & WHT/UCAM & Clear, $1.1$''; $u'$, $g'$, $r'$.\\ 
\hline
\end{tabular}
\end{minipage}
\end{table*}
The majority of data were acquired with the high-speed cameras ULTRACAM
\citep{Dhillon2007MNRAS.378.825} and RISE
\citep{Steele2008SPIE7014.217,Gibson2008AA492.603}. These employ
frame transfer CCDs so that deadtime between images is reduced to less than
$0.05$ seconds.  ULTRACAM, a visitor instrument, was mounted either at a
Nasmyth focus of the 3.5m New Technology Telescope (NTT) in La Silla or the
Cassegrain focus of the 4.2m William Herschel Telescope (WHT) in La Palma,
while RISE is permanently mounted on the robotic 2m Liverpool Telescope
(LT). The robotic nature of the LT allows us to spread the observations,
while ULTRACAM provides the highest precision data. We used $u'$ and 
$g'$ filters in the blue and green channels of ULTRACAM and $r'$ or $i'$ in
the red arm, as listed in Table~\ref{tab:obslog}. RISE operates with a single 
fixed filter spanning the $V$ and $R$ bands. We also observed NN~Ser 
with the infrared imager HAWK-I installed at the Nasmyth focus of VLT-UT4 at 
Paranal \citep{kissler08} in March and April 2013. We used
the fast photometry mode which allowed us to window the detectors and achieve
a negligible dead time between frames. Observations were performed using the
$J$-band filter; the white dwarf contributes $\sim$60\% of the overall
light in this band meaning that the eclipse is still deep and suitable for
timing. 

All data were flat-fielded and extracted using aperture photometry within the
ULTRACAM reduction pipeline \citep{Dhillon2007MNRAS.378.825}. We fitted the
resulting light curves using the light curve model developed in our previous
analysis of NN~Ser \citep{Parsons2010MNRAS402.2591}. Holding all parameters
fixed except the eclipse time led to the measurements listed in
Table~\ref{tab:obslog}, with the uncertainties derived from the covariance matrix
returned from the Levenberg-Marquardt minimisation used. In each case we
scaled the uncertainties on the data to ensure a $\chi^2$ per degree of
freedom equal to one. We estimate uncertainties on our data by propagation of
photon and readout noise through the data reduction. In good conditions these
give realistic estimates of the true scatter in the data, and the scaling
therefore makes little difference. In poor conditions the scatter can be
larger than the error propagation suggests in which case the scaling returns
larger, more realistic uncertainties. It is changes in the observing
conditions, as well as the instruments, that largely account for the variation
in the uncertainties listed in Table~\ref{tab:obslog}, with the addition of pickup
noise that affected ULTRACAM in January 2012 owing to a faulty data cable.  In
the case of the ULTRACAM data, we combined the times from the three
independent arms of ULTRACAM, weighting inversely with variance to arrive at
the times listed. The first two times listed in Table~\ref{tab:obslog} represent
updates of times listed in \citet{Beuermann2010AA521L60} which were based upon
the $g'$-arm of ULTRACAM only; the remainder of the times we used are as
listed in \citet{Beuermann2010AA521L60}. Adding our data to those of
\citet{Beuermann2010AA521L60} gives a total of 76 times.  One
  eclipse listed in Table~\ref{tab:obslog}, that of cycle 66905, was very badly
  affected by cloud on both ingress and egress ($> 90$\% and $\sim 50$\% loss
  of light). During egress, the cloud was thinning, leading to a rising trend
  in throughput which weights the flux towards the second half of each
  exposure, and can be expected to delay the measured time. Consistent with
  this, the time for this eclipse is significantly delayed with respect to the
  best fit, and including it in the fits adds $14.5$ to $\chi^2$. We therefore
  decided to exclude it from the analysis of the paper, but list it in
  Table~\ref{tab:obslog} for completeness.

For timing, precision is largely a matter of telescope aperture and noise
control; accuracy is down to the data acquisition system and the corrections
needed to place the times onto a uniform scale. Significant timing errors have
been found in the data of \citet{Dai2010MNRAS409.1195} for UZ~For, and in the
data of \citet{Qian2011MNRAS414L16} for HU~Aqr
\citep{Potter2011MNRAS416.2202,Gozdziewski2012MNRAS425.930}, and these are
just ones that have been spotted from independent work, thus attention must
always be paid to the absolute timing accuracy of instruments in such
work. For ULTRACAM we have measured the absolute timing to be good to $<
0.001\,\sec$; RISE is measured to be good to better than $0.1\,\sec$
(Pollacco, priv.\ comm.).  While this upper limit potentially allows
systematic errors which are larger than the smallest uncertainties from
ULTRACAM timing of NN~Ser, it is below the uncertainties of times based upon
RISE itself. In HAWK-I's fast photometry mode data is collected in blocks of
exposures. There is an overhead between blocks of 1--2 seconds as the data are
written to disk. Only the first exposure of each block is timestamped (to an
accuracy of $\sim$10 milliseconds) therefore we used a small block size of 30
exposures in order to reduce the timing uncertainties on the subsequent
exposures within a block. Since the dead time between exposures within a block
is negligible, we estimate that the timing accuracy of HAWK-I is better than
0.1 seconds, smaller than the uncertainties on the eclipse times measured with
HAWK-I.

The times were placed on a TDB (Barycentric Dynamical Time) timescale
corrected for light travel effects to the barycentre of the solar system to
eliminate the effect of the motion of Earth (see \cite{Eastman2010PASP122.935}
for more details of time systems).  We carried out these corrections with a
code based upon SLALIB, which we have found to be accurate at a level of $50$
microseconds when compared to the pulsar timing package TEMPO2
\citep{Hobbs2006MNRAS369.655}, an insignificant error compared to the
statistical uncertainties of our observations. We quote the times in the form
of modified Julian dates, where $\mathrm{MJD} = \mathrm{JD} - 2400000.5$,
because this is how we store times for increased precision. Placed upon a TDB
timescale this becomes MJD(TDB), and it takes its final form BMJD(TDB) when
corrected to the barycentre of the solar system.

\section{Analysis and Results}

We begin our presentation of the results with two sections outlining 
the analysis methods we used. The second of these concerns the numerical
aspects of fitting models to data, while we start with a discussion of the
physical models adopted.

\subsection{Description of the orbits}
\label{sec:kepnew}

We assume the binary acts as a clock which moves relative to the observer
under the influence of unseen bodies, hereafter ``planets'', in bound orbits
around the binary. Labelling the binary with index $0$ and the planets with
indices $1$, $2$, \ldots $N$, we need to describe the orbits of $N+1$
bodies. The most direct method is to specify the Cartesian coordinates and
velocities of the $N+1$ bodies at a given time, $6(N+1)$ parameters in all. By
working in the barycentric (centre-of-mass) frame, this can be reduced to $6
N$ without loss of generality. We use the $6 N$ parameters to specify the
barycentric positions $\vec{R}_i$ and velocities $\vec{V}_i$, $i = 1 \ldots N$,
of the $N$ planets at a specific time, with the binary's position and velocity
determined through the reflex condition
\begin{equation}
m_0 \vec{R}_0 = - \sum_{i=1}^N m_i \vec{R}_i, \label{eq:reflex}
\end{equation}
where $m_0$ and $m_i$ are the masses of the objects, with a similar condition
on the velocity. This is how we initialise our $N$-body integrations, which we
will describe later.

For two-body orbits it is more usual to characterise orbits in terms of six
Keplerian orbital elements ($a$, $e$, $i$, $\Omega$, $\omega$, $T_0$, to be
defined later) together with Kepler's third law which gives the orbital
angular frequency in terms of the masses of the bodies and semi-major axis of
the orbit. For two-body orbits, Keplerian elements are time-independent,
unlike the Cartesian vectors.  In trying to extend them to the case of more
than one planet ($N > 1$), we face two problems. First, when there are more
than two bodies, Keplerian orbits are only an approximation to the true,
hereafter Newtonian, orbits and we need to determine whether the degree of
approximation is significant. Second, there is more than one way to
parameterise the orbits in terms of Keplerian motion, and each differs in
terms of how well it approximates the Newtonian paths.

We consider three alternative orbit parameterisations. The first two have
already appeared in the literature, while the third, which has not been
presented before as far as we are aware, performed better than the other
two. The three parameterisations differ in how we define the vectors which
undergo Keplerian motion and in the precise forms of Kepler's third law that
we use.

We call our first parameterisation ``astrocentric''. The coordinates of each
planet are referenced relative to the binary, and we
assume that each astrocentric vector follows its own Keplerian two-body orbit,
with angular frequencies $n_i$ given by
\begin{equation}
n_i^2 a_i^3 = G (m_0+m_i),
\end{equation}
for $i = 1, \ldots, N$. These are the coordinates used when fitting eclipse
times by most researchers to date. In astrocentric coordinates each planet is
placed upon the same footing, and is treated as if the other planets were not
there. Denoting astrocentric vectors by the lowercase greek letter
$\vec{\rho}$, the position vector $\vec{\rho}_0$ points from the barycentre of
all the bodies to the binary, and then the vectors $\vec{\rho}_i$ point from
the binary to the planets. In astrocentric coordinates the reflex condition
Eq.~\ref{eq:reflex} becomes
\begin{equation}
\vec{\rho}_0  = - \sum_{i=1}^N k_i \vec{\rho}_i, \label{eq:zero}
\end{equation}
where $k_i = m_i / M$, where $M = \sum_{j=0}^N m_j$ is the total mass.  We
will encounter these parameters in slightly modified form for the other two
parameterisations. A typical procedure is to start with $N$ sets of Keplerian
elements from which the $N$ vectors $\vec{\rho}_i$, $i = 1, \ldots, N$ can be
calculated. The binary vector $\vec{\rho}_0$ then follows from
Eq.~\ref{eq:zero}, and the equivalent barycentric vectors follow from
\begin{equation}
\vec{R}_i  = \vec{\rho}_i  + \vec{\rho}_0 . \label{eq:general}
\end{equation}

Despite their simplicity, astrocentric coordinates are unattractive from a
theoretical point of view. If one transforms from barycentric to astrocentric
coordinates, the kinetic energy part of the Hamiltonian, which in barycentric
coordinates is
\begin{equation}
 H_K = \frac{1}{2} \sum_{i=0}^N m_i \dot{\vec{R}}_i^2,
\end{equation}
develops cross-terms such as $\dot{\vec{\rho}}_1 \dot{\vec{\rho}}_2$. This
problem can be avoided using Jacobi coordinates \citep{Malhotra1993ASPC36.89M},
and orbits prove to be closer to Keplerian in these coordinates than they do in
astrocentric coordinates \citep{Lee:2003ApJ592.1201}; this was first pointed
out for planets around white dwarf binaries by
\cite{Gozdziewski2012MNRAS425.930}.  We use Jacobi coordinates for the second
and third parameterisations as we now discuss.

Jacobi coordinates, which we indicate with lowercase latin letter $\vec{r}$, are
defined as follows: vector $\vec{r}_0$ points from the system barycentre to
the binary; $\vec{r}_1$ points from the binary to the first planet;
$\vec{r}_2$ points from the centre of mass of the binary and first planet
towards the second planet, and so on, with each new vector pointing from the
centre of mass of the combined set of objects up to that point to the next
object. These coordinates differ from the astrocentric series $\vec{\rho}_0$,
$\vec{\rho}_1$, $\vec{\rho}_2$, \ldots, only from the third term onwards, and
are therefore no different in the two body case. It can be shown
\citep{Malhotra1993ASPC36.89M} that in Jacobi coordinates the kinetic energy
part of the Hamiltonian takes the simple form
\begin{equation}
 H_K = \frac{1}{2} \sum_{i=1}^N \mu_i \dot{\vec{r}}_i^2,
\end{equation}
where $\mu_i$ is the reduced mass of planet $i$ in orbit with a single
object consisting of the binary and all planets up to number $i-1$:
\begin{equation}
 \frac{1}{\mu_i} = \frac{1}{\sum_{j=0}^{i-1} m_j} + \frac{1}{m_i} .
\end{equation}
For three bodies the overall Hamiltonian can then be written as
\begin{equation}
 H = \sum_{i=1}^2 \left(\frac{1}{2} \mu_i \dot{\vec{r}}_i^2 - \frac{G m_0
    m_i}{r_i}\right)  + H',
\end{equation}
where 
\begin{equation}
 H' = Gm_0 m_2 \left(\frac{1}{r_2} -
    \frac{1}{|\vec{r}_2 + k'_1 \vec{r}_1|} \right) - \frac{G m_1
    m_2}{|\vec{r}_2-(1-k'_1)\vec{r}_1|},\label{eq:hint}
\end{equation} 
and $k'_1$ is one of a series of factors related to the centre-of-mass
sequence:
\begin{equation}
k'_i = \frac{m_i}{\sum_{j=0}^i m_j} , \;\; i = 1, 2, \ldots N.
\end{equation}
Since $k'_1 = m_1/(m_0+m_1)$, both terms in Eq.~\ref{eq:hint} are of order
$m_1 m_2$ \citep{Malhotra1993ASPC36.89M}. If the planet masses are very small
compared to $m_0$, we can neglect $H'$ with respect to the terms of the
summation, and the problem simplifies to two Kepler orbits in the Jacobi
coordinates for each planet, $\vec{r}_1$ and $\vec{r}_2$, with orbital angular
frequencies $n_1$ and $n_2$ given by
\begin{eqnarray}
n_1^2 a_1^3 &=& G \frac{m_0}{1-k'_1} = G (m_0+m_1), \label{eq:first}\\
n_2^2 a_2^3 &=& G \frac{m_0}{1-k'_2} = G \frac{m_0 (m_0+m_1+m_2)}{m_0 + m_1} \label{eq:second}.
\end{eqnarray}
The factors $k'_i$ are analogous to the $k_i$ introduced for astrocentric
coordinates, and appear in the following relations that correspond to
Eqs~\ref{eq:zero} and \ref{eq:general}:
\begin{equation}
\vec{r}_0  = - \sum_{i=1}^N k'_i \vec{r}_i,
\end{equation}
and
\begin{equation}
\vec{R}_i  = \vec{r}_i  - \sum_{j=i}^N k'_j \vec{r}_j .
\end{equation}

Eq.~\ref{eq:second} relating the orbital frequency $n_2$ to the semi-major
axis $a_2$, is slightly unexpected. The form of the reduced mass $\mu_2$
suggests that this should represent a composite object consisting of the
binary and first planet with total mass $m_0 + m_1$, in orbit with the second
planet of mass $m_2$. Hence one might have guessed that Eq.~\ref{eq:second}
would simply read $G (m_0+m_1+m_2)$ on the right-hand side.  This is the
motivation behind our third and final set of coordinates, which, for want of a
better term, we name ``modified Jacobi coordinates''. The only change we make
for the modified Jacobi coordinates is to alter Eq.~\ref{eq:second} to read
\begin{equation}
 n_2^2 a_2^3 = G (m_0+m_1+m_2).
\end{equation}
This choice corresponds to a slightly different partitioning of the
Hamiltonian in which the perturbation Hamiltonian takes on the 
modified form
\begin{eqnarray}
 H'' &=& G m_0 m_2 \left(\frac{1}{r_2} -
    \frac{1}{|\vec{r}_2 + k_1 \vec{r}_1|} \right) + \nonumber \\
  && G m_1 m_2 \left(\frac{1}{r_2} -
    \frac{1}{|\vec{r}_2-(1-k_1)\vec{r}_1|}\right).
\end{eqnarray}
Just as for $H'$, both terms are of order $m_1 m_2$, but $H''$ is better for a truly
hierarchical set of orbits since if $r_1 \ll r_2$, the second term is much
smaller than it is in $H'$.

In contrast to the astrocentric case, the two planets are not treated
symmetrically by Jacobi coordinates and thus their ordering
matters. Considering $H''$, the order-of-magnitude of both terms is $Gm_1 m_2
r_1/r_2^2$, thus the correct choice is to label the planets so that $r_2 >
r_1$, i.e. planet~1 should be the closest to the binary. This reduces the size
of $H''$ by the ratio of the semi-major axes squared, $\sim (a_1/a_2)^2$,
relative to the reverse choice.  Hence in the rest of the paper, we number the
planets in ascending order of their semi-major axes, with planet~1 the
innermost.

We have emphasised that Keplerian orbits are an approximation for $N >
1$. However, Keplerian elements can simply be regarded as a set of generalised
coordinates which vary with time for $N > 1$. Such ``osculating'' elements
precisely specify the paths of the bodies, although the way in which the
elements evolve with time must be determined through numerical $N$-body
integration. Each of the three parameterisations can be used in this way, as
well as in the Keplerian approximation with all elements fixed. To do so one
starts from a set of elements at a particular time, which are then translated
into barycentric Cartesian coordinates. One then proceeds using $N$-body
integration thereafter. The translation step varies with the parameterisation
in use, so identical $N$-body paths correspond to slightly different sets of
elements according to the chosen parameterisation, but used in this way the
orbits are exact within numerical error, which allows us to judge the degree
of approximation involved in the Keplerian approximation.

We wrote a numerical $N$-body integrator in C++ based upon the Burlisch-Stoer
method as implemented by \cite{Press2002NR}, which we ran from within a Python
wrapper. We verified our integrator on the Kepler 2-body problem, an
equal-mass symmetric three-body problem, against an entirely independent code
written by one of us (MB), and against the Burlisch-Stoer option of the orbit
integrator, MERCURY6 \citep{Chambers1999MNRAS304.793}. For each of the three
parameterisations we computed $N$-body-integrated paths to equivalent 
Keplerian approximated orbits. We selected $\mathrm{MJD} = 54500$,
which corresponds to Feb 4, 2008 as the reference epoch since it is weighted
towards the era when the bulk of high quality eclipse times have been taken.
We verified the significance of the planet ordering for the two forms of
Jacobi coordinates, finding that the correct choice was better than the
reverse by of order a factor of 5 in terms of RMS difference versus Newtonian
models.

\begin{figure}
\hspace*{\fill}
\includegraphics[width=0.97\columnwidth]{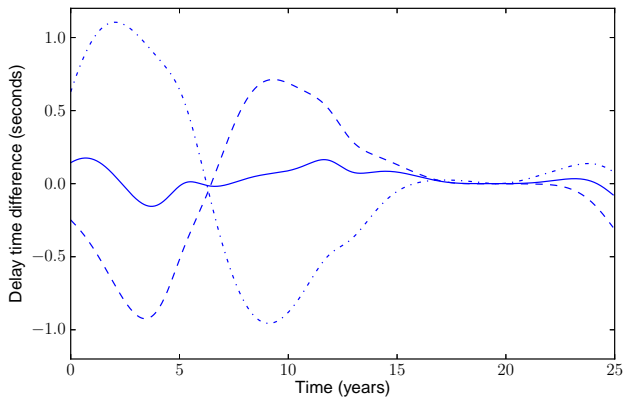}
\hspace*{\fill}

\caption{The difference in the planet-induced light-travel-time (LTT) delays
  of Keplerian versus Newtonian models for a typical two-planet fit for
  NN~Ser. Three cases are shown: astrocentric (dashed-dotted), Jacobi (dashed)
  and modified Jacobi (solid). The $\mathrm{MJD} = 54500$ reference time
  corresponds to the time around 19 years in when all models agree. For
  reference the LTT variations in NN~Ser have a range of $\pm
  40\,$seconds. The Newtonian comparison orbits are calculated separately for
  each of the three coordinate systems.
\label{fig:compare_coord}}
\end{figure}

Fig.~\ref{fig:compare_coord}
compares the difference between Keplerian and Newtonian predictions for the
three parameterisations for an orbit typical of NN~Ser. The ordering seen here
with astrocentric coordinates worst, and our modified version of Jacobi
coordinates best, agrees with what we found looking at a much broader range of
orbit fits. The differences in Fig.~\ref{fig:compare_coord} range from a few
tenths of a second to more than one second, which given the timing precision
of NN~Ser can be expected to have a noticeable effect upon parameters. There
are instances where deviations as large as 5 seconds occur, typically on
dynamically very unstable orbits. We will see that these can have a
quantitative effect upon the parameters, meaning that Keplerian models,
whatever the coordinate parameterisation, are not adequate for fitting the
NN~Ser times.  In consequence, the majority of the orbit fits in this paper,
were undertaken using Newtonian $N$-body integrations, without Keplerian approximation. We employed the
modified Jacobi representation to translate from orbital elements to initial
position and velocity vectors to initialise these integrations, because, as
Fig.~\ref{fig:compare_coord} shows, they are the best of the three we
investigated.  We make one exception where we compare the results from
$N$-body integrated and equivalent Keplerian models, based in each case upon
the modified Jacobi prescription. When we need to specify exactly what system
we are using, we will use expressions such as ``astrocentric Keplerian'' and
``Newtonian modified Jacobi''. The first means orbits in which two
astrocentric vectors execute Kepler ellipses, i.e. an approximation; the
second means that Jacobi coordinates are used to initialise the orbits, using
our modified version of angular frequency, but thereafter the paths are
computed using $N$-body integration with no approximation beyond numerical
uncertainties.

\begin{figure*}
\includegraphics[angle=270,width=0.8\textwidth]{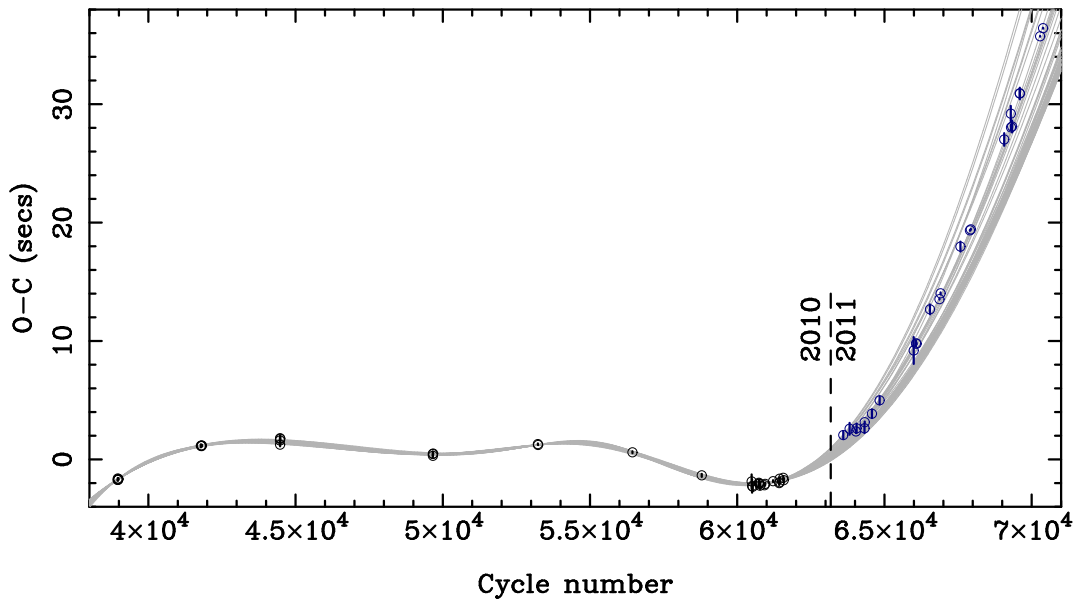}
\caption{Eleven years of eclipse times of NN~Ser, starting in May 2002. The
  dashed line marks the end of 2010; the data before this point are those
  listed in \protect\citet{Beuermann2010AA521L60}, including the two updates
  listed in Table~\protect\ref{tab:obslog}. The times are plotted relative to
  the ephemeris $\mathrm{BMJD(TDB)} = 47344.0258547 + 0.1300801135 E$, where
  $E$ is the cycle number. This was chosen to give a flat trend in times from
  2002 to 2010.  The light-grey smooth curves show 50 Newtonian orbit fits to
  the pre-2011 data only, generated via MCMC iteration, corresponding to the
  models of the lower-left panel of Fig.~\protect\ref{fig:p1p2}; the models
  were picked from the final 100 models of the MCMC chain. The times after
  2010 are from this paper and were not used to create the fits, and yet they
  match the predicted trend well. For clarity, only data with
    uncertainties $< 2\,$sec are shown. \label{fig:predict}}
\end{figure*}

\subsection{Model fitting approach}
\label{sec:mcmc}
Sometimes-sparse coverage, and often-long orbital periods, mean that timing
work on circum-binary planets is plagued by degeneracies amongst fit 
parameters. This can cause problems simply in locating best-fit models, and
even more so in the determination of uncertainties. For instance the
widely-used Levenberg-Marquardt method often fails to locate the minimum in such
circumstances and the covariance matrix it generates can be far from capturing
the complexity of very non-quadratic, and possibly multiple minima. A
widely-used method that can overcome these difficulties, which we
adopt here, is the Markov Chain Monte Carlo (MCMC) method. The aim of MCMC
analysis is to obtain a set of possible models distributed over model
parameter space with the Bayesian posterior probability distribution 
defined by the data. This is accomplished by stochastic jumping of the model
parameters, followed by selection or rejection according to the posterior
probability of the model $M$ given the data $D$, $P(M|D)$. This process
results in long chains of models, which, if long enough to be well-mixed, have
the desired probability distribution. By Bayes' theorem the posterior
probability is proportional to the product of the prior probability of the
model, $P(M)$, and the likelihood, $P(D|M)$, which in our case is
determined by the factor $\exp (- \chi^2/2 )$, where $\chi^2$ is the standard
goodness-of-fit parameter.

For the prior probabilities, we adopted uniform priors for all temporal
zero-points, the eccentricities (0 to 1), and the arguments of periapsis
($-180^\circ$ to $+180^\circ$). We used Jeffreys priors ($1/a$, $1/m$) for the
semi-major axes and masses. Some care is needed over the eccentricity $e$ and
the argument of periapsis $\omega$, which sets the orientation of the ellipse
in its own plane, because $\omega$ becomes poorly constrained as $e
\rightarrow 0$.  This can cause difficulties if one iterates using $e$ and
$\omega$ directly. We therefore transformed to $x = \sqrt{e} \cos \omega$ and
$y = \sqrt{e} \sin \omega$, which since the Jacobian
$||\partial(x,y)/\partial(e,\omega)||$ is constant, maintains uniform
priors in $e$ and $\omega$, but causes no difficulties for
small values of $e$. The choice of priors has a small but non-negligible
effect upon the results. For instance we find a significant range of
semi-major axes in some models, and there is clearly a difference between a
uniform prior and $1/a$. Although the priors can have a quantitative effect
upon results in such cases, they have no qualitative impact upon the
conclusions of this paper.

Armed with the MCMC runs, we are in a position to compute uncertainties, and
correlations between parameters. The MCMC method is useful in cases of high
dimensionality such as we face here (the models we present require from 10 to
13 fit parameters) and can give a good feel for the regions of parameter space
supported by the data. Requiring no derivative information, it is highly
robust, a significant point for the Newtonian models where one can generate
trial orbits which do not even last the span of the observed data. These cause
difficulties for derivative-based methods such as Levenburg-Marquardt for
example.  Generation of models with the correct posterior probability
distribution is also ideal for subsequent dynamical analysis where one wants
to tests models that are consistent with the data.

The main disadvantage of the MCMC method is the sometimes-large computation
time needed to achieve well-mixed and converged chains. The way in which the
models are jumped during the iterations is important. Small jumps lead to slow
random-walk behaviour with long correlation times, while large jumps lead to a
high chance of rejection for proposed models and long correlation times once
more. Ideally one jumps with a distribution that reflects the correlations
between parameters, but it is not always easy to work out how to do this, and
there is no magic bullet to solve this in all cases. For instance if multiple
minima are separated by high enough $\chi^2$ ``mountains'', a chain may never
jump between them. In this paper we adopted the affine-invariant method
implemented in the Python package \texttt{emcee}
\citep{Foreman-Mackey:2013PASP125.306}. This adapts its jumps to the
developing distribution of models, which is a great advantage over having to
estimate this at the start, but even so, the problem in this case turned out
to be one of the most difficult we have encountered, and in several cases we
required $> 10^9$ orbits to reach near-ergodic behaviour. We computed the
autocorrelation functions of parameters as one means of assessing convergence,
but our main method, and the one we trust above any other, was visual, by
making plots of the mean and root-mean-square (RMS) values of parameters as a
function of update cycle number along the chains. Initial ``burn-in'' sections
are obvious on such plots, as are long-term trends. There is no way to be
absolutely certain that convergence has been reached in MCMC because there can
be regions of parameter space that barely mix with each other. Even if one
computed $10^{10}$ models, there would be no guarantee that a new region of
viable models would not show up after $10^{12}$. From the very many
computations we have carried out, including large numbers of false starts, we
believe that we have explored parameter space very fully, and there are no
undiscovered continents of lower $\chi^2$. However, as we will
describe later, we did encounter one case that converged too slowly to give 
reliable results. This is fundamentally an issue of degeneracy and it should 
improve greatly with further coverage.

\subsection{Predicting the future}
We start our analysis with our primary objective: how well does the two-planet
model developed by \citet{Beuermann2010AA521L60} fare when confronted with new
data?  Fig.~\ref{fig:predict} shows the most recent eleven years of data on
NN~Ser, dating back to May 2002 when we first started to monitor it with
ULTRACAM. The vertical dashed line at the end of 2010 marks the boundary
between the times listed in \citet{Beuermann2010AA521L60} and the new times of
this paper. The grey curves are a sub-set of 50 MCMC-generated Newtonian
models \textit{based upon \citet{Beuermann2010AA521L60}'s times
  alone}. Without the new times or orbit fits to guide the eye, one might have
guessed that the new times would perhaps range in $O-C$ around $\pm 3\,\sec$
on this plot. However earlier data, which are included in the fits, but off
the left-hand side of the plot windows (see \citet{Beuermann2010AA521L60} and
Fig.~\ref{fig:future} later in this paper), cause the planet model to predict
a sharp upturn since 2010, corresponding to delayed eclipse times as the binary
moves away from us relative to its mean motion during the previous 8 years. In
the planetary model, the upturn is primarily the result of the $7\,\mj$
outermost planet. Our new data are in remarkably good agreement with this
(remarkable to the authors at least). While this is not a proof of the
planetary model, it has nevertheless passed the test well. We can't say for
sure that alternative models such as those of \citet{Applegate1992ApJ385.621}
don't have a similarly precise ``memory'' of the past, but neither is it clear
that they do, whereas it is a key prediction of the clockwork precision of
Newtonian dynamics.

\begin{figure}
\includegraphics[angle=270,width=\columnwidth]{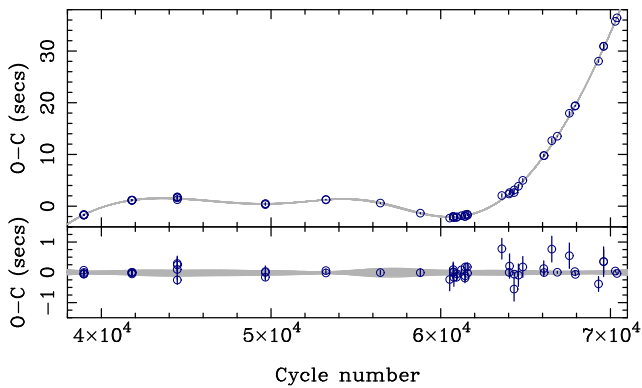}
\caption{This plot is identical to Fig.~\protect\ref{fig:predict} but now the
  orbital fits are based upon all data, incorporating the new times, and it
  includes a plot of the residuals relative to the best of the orbits
  shown. For clarity, only points with uncertainties $< 0.5\,$sec are shown.
\label{fig:refined}}
\end{figure}

Including the new times when generating the fits, gives a much tighter set of
possible orbits illustrated in Fig.~\ref{fig:refined}
\begin{figure*}
\includegraphics[width=0.8\textwidth]{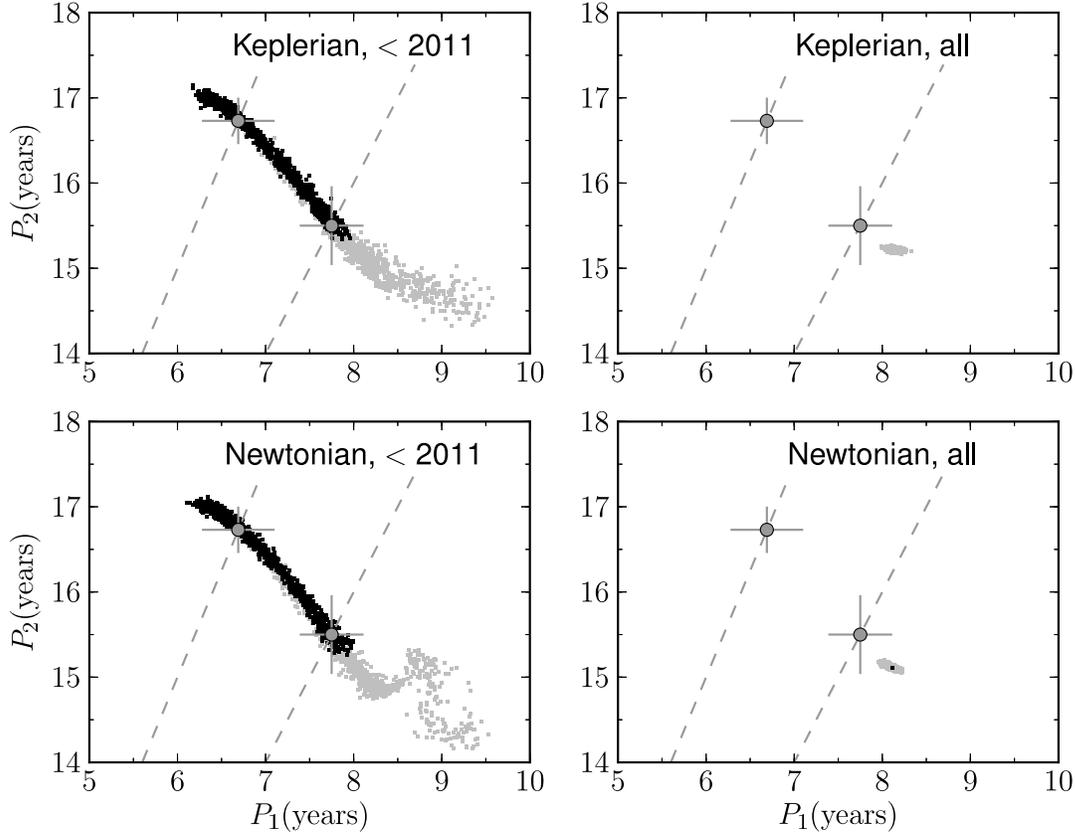}
\caption{Regions of $P_1$--$P_2$ space supported by the data, shown
  in each case using 2000 models sampled from MCMC chains. The top-left panel
  is our equivalent of \protect\citet{Beuermann2010AA521L60}, i.e. we use only
  data taken before 2011 and assume Keplerian orbits (although our Jacobi
  formalism leads to a very small change in position).  The top-right panel
  shows Keplerian models based upon all of the data; the lower panels show the
  corresponding results for Newtonian orbit integrations. The grey dashed
  lines mark 2:1 (right-hand) and 5:2 (left-hand) period
  ratios, while the crosses mark the models ``2a'' (lower right) and
  ``2b'' (upper left) from \protect\citet{Beuermann2010AA521L60}.
  Black (grey) points delineate models which last either more (less) than 1
  million years, the post-common-envelope age of NN~Ser.
\label{fig:p1p2}}
\end{figure*}
which also shows residuals between the data and the best of the fits shown. It
should be noted however that at this point we are following
\citet{Beuermann2010AA521L60}'s assumption of zero eccentricity for the outer
orbit, which is largely responsible for the very tightly defined fit. The
dispersion increases once this constraint is lifted (independent of whether
Newtonian or Keplerian models are adopted).

\subsection{Comparison with \protect\citet{Beuermann2010AA521L60}}

The fits plotted in Figs~\ref{fig:predict} and \ref{fig:refined} were based
upon allowing the same parameters to vary as used in
\citet{Beuermann2010AA521L60}'s model~''2a'' (their best one), so in this
section we look at the effect that the new data has upon the parameters. We
also consider the difference made by using integrated Newtonian models
compared to Keplerian orbits; in all subsequent sections we use Newtonian
models only. For reference, in their (astrocentric Keplerian) model~2a,
\citet{Beuermann2010AA521L60} allowed a total of 10 parameters to be free
which were the zero-point and period of the binary's ephemeris, the period,
semi-major axis and reference epoch of the outer planet, and the period, semi-major
axis, reference epoch, eccentricity and argument of periastron of 
inner and lower mass planet. The orbit of the outer planet
was assumed to be circular.

\citet{Beuermann2010AA521L60} give a detailed description of their fits in
terms of the periods ``$P_c$'' and ``$P_d$'' of the two planets
(corresponding to our $P_2$ and $P_1$), so we first focus upon
this. Fig.~\ref{fig:p1p2} shows the range of $P_1$--$P_2$ space
supported under either the Keplerian or Newtonian interpretations, and making
use of either the data used by \citet{Beuermann2010AA521L60} only, or the full
set including our new times. The top-left panel is equivalent to
\citet{Beuermann2010AA521L60} and indeed matches the range of models they
located, although the MCMC results show that the supported region is more
complex than their division into just two models perhaps suggests. The
top-right panel shows a significant shrinkage with the addition of new data
and supports \citet{Beuermann2010AA521L60}'s selection of their
model~2a. While some shrinkage is expected, the extent of the change is
notable, given that we have have only increased the baseline of coverage by
around 15\%. We believe this is a combination of degeneracy when fitting to
pre-2011 data alone, combined with our having turned the corner of another
orbit of the outer planet (planet 2), as shown by Fig.~\ref{fig:predict}.
\citet{Beuermann2010AA521L60} found that there is little to choose between
their two models in terms of goodness of fit, although their model~2a was
marginally favoured. This is confirmed by the stripe of viable models
connecting their 2a and 2b in the top-left panel of Fig.~\ref{fig:p1p2}.

The lower panels show that, even though our choice of coordinates was
motivated by the desire to generate Keplerian orbits which matched Newtonian
orbits as closely as possible, there are nonetheless regions of parameter
space considerably affected by three-body effects. In particular, the kink in
the lower-left panel located in the region where the period ratio is closer
than 2:1, compared to its relatively simple Keplerian counterpart in the
upper-left panel, is evidence of this. Here deviations between Keplerian and
Newtonian orbits amount to several seconds, highly significant given the
precision of the NN~Ser times, and the favoured parameter distribution is
distorted as a result. The effects are much smaller above the 2:1
line, and show that the modified Jacobi coordinates can work well.  Strangely
enough, as we remarked earlier, although three-body effects are significant,
the data are not good enough to prove that they operate (which could provide
compelling independent support for the planet model) because there is
sufficient degeneracy for either Keplerian or Newtonian models to fit the data
equally well, albeit with differing sets of orbital elements. Obviously, if
there are planets orbiting the binary in NN~Ser, the weight of 300 years of
classical mechanics favours Newtonian models, but it will be some time before
this can be proved from the data directly.

\subsection{Dynamical stability}

As discussed earlier, some proposed circum-binary orbits have been shown to be
unstable on short timescales, and if multiple planetary orbits are proposed, a
check on their stability is essential. Having said this, all the data needed
for this are not to hand since we don't know the mutual orientations of the
planets' orbits. Therefore, in the absence of evidence to the contrary, we
assume, along with previous researchers, that the orbits are coplanar. In
addition we assume that, like the binary itself, we see the planetary orbits
edge-on and for simplicity we set the orbital inclinations precisely to
$90^\circ$. This minimises the masses of the planets relative to the binary,
which will usually tend to promote stability. NN~Ser emerged from its common
envelope phase around one million years ago, and prior to this phase would
have been significantly different, so we checked for stability by integrating
backwards in time for just 2 million years. To a certain extent stability is
already included within the Newtonian MCMC runs (lower panels of
Fig.~\ref{fig:p1p2}) since some proposed orbits generated by MCMC jumps lead
to collisions within the span of the data and are rejected. It would have been
easy to extend this so that all long-term unstable orbits were similarly
thrown out, however, the CPU time penalty is far too great to allow this
approach. Instead, our approach during the MCMC runs was simply to integrate
for the 25 year baseline of the observations, leaving the longer-term
dynamical stability computations to the small fraction of models retained (of
order 1 in $10^4$) as we waited for the MCMC chains to reach a stable state.

\begin{figure*}
\includegraphics[width=\textwidth]{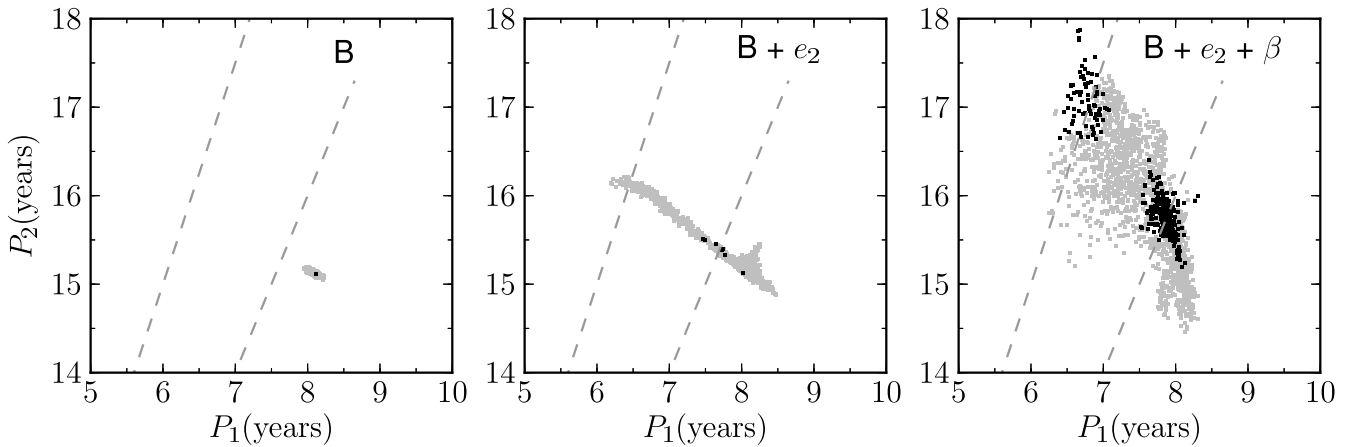}
\caption{Regions of $P_1$--$P_2$ space supported by the data, showing the
  change as the models are given greater freedom. The left-most panel is the
  constrained model~2 (``B'' for short) from
  \protect\citet{Beuermann2010AA521L60} for reference (i.e. it is
  identical to the lower-right panel of
  Fig.~\protect\ref{fig:p1p2}). In the centre panel, the eccentricity of the
  outermost planet is allowed to be non-zero, while the right-most panel
  allows the binary's period to change in addition. Each panel shows 2000
  Newtonian models based upon all of the data. As before, the grey dashed
  lines mark 2:1 (top) and 5:2 (bottom) period ratios, and black and grey
  points indicate stable and unstable models. From left-to-right, 0.02\%,
  0.7\% and 15\% of the models last more than 1 million
  years.\label{fig:p1p2_eq}}
\end{figure*}

The differently shaded symbols in Fig.~\ref{fig:p1p2} distinguish between
``stable'' orbits which last for $> 1$ million years (black) from the
``unstable'' ones which do not (grey).  In the upper-left panel,
orbits are mostly unstable below the 2:1 line (i.e. less extreme ratio), as
one might expect. They are stabilised near the 2:1 and 5:2 lines and there is
a mixed set of unstable and stable models in between. The pattern of stability
and instability is broadly consistent with \citet{Beuermann2010AA521L60}'s
results, although our models seem to be more stable between the 2:1 and 5:2 lines
than their description suggests. The topology of stable and unstable
regions survives the distorting influence of Newtonian effects in the
lower-left panel. Of order 50\% of these models proved to be stable. Once the
new data are included (right-hand panels), the supported models are confined
to the largely unstable region lying below the 2:1 line in
Fig.~\ref{fig:p1p2}. Unsurprisingly therefore, very few of these models turn
out to be stable -- around 1 in 6000. Although one could argue that just one
stable model consistent with the data is all that is required to claim
potential stability, the reduction in the fraction of stable models is a worry
for the planet model of NN~Ser, because it looks possible that with yet more
data, we are likely to be left with no long-lived models at all. Thus we now
turn to look at the consequences of freeing up the orbit fits by allowing
non-zero eccentricity in the outermost planetary orbit and changes in the
orbital period of the binary itself.

\subsection{Eccentricity and binary orbital period variation}

We have so far followed \citet{Beuermann2010AA521L60}'s application of
Ockham's razor by choosing the most restrictive model consistent with the
data. This helps the fitting process because of degeneracies, as
\citet{Beuermann2010AA521L60} suggest, but it gives an overly optimistic view
of how well constrained NN~Ser is. In following
\citet{Beuermann2010AA521L60}'s model~2, we are making the questionable
assumptions that the outer planet has a circular orbit and that
NN~Ser acts as a perfect clock. While we don't need to deviate from
these in order to find good fits to the data, it would come as little surprise
if they were not entirely accurate, so it is of interest to examine the effect
relaxing these restrictions has upon the model parameters, and also upon the
issue of stability, which, as we have just seen, is looking marginal in the
light of the new data.  We therefore carried out MCMC runs with the outermost
planet's orbit allowed to be eccentric (two extra free parameters, making 12),
and then with the addition of ``clock drift'' in the form of a quadratic term
$\beta$ in the binary ephemeris, bringing the number of free parameters to
13. We found that the MCMC chain of the last case never converged
owing to a strong degeneracy between the quadratic term and the orbit of
the outer planet which allowed $a_2$ to range up to values $> 12\,$AU
compared to a value $\approx 5.4\,$AU when no quadratic term was
included. In order to force convergence upon a reasonable timescale, we
therefore applied a gaussian prior on $\beta$, where the latter is defined
by its appearance in the ephemeris relation
\begin{equation}
T = T_0 + P E + \beta E^2,
\end{equation}
with $E$ the eclipse cycle number and $T$ the time in days. The prior we
applied was $P(\beta) \propto \exp (-(\beta/\sigma_\beta)^2/2)$, with
$\sigma_\beta = 7.5 \times 10^{-14}\,$days, 25 times the magnitude expected
for gravitational wave losses (see later). This allows significant extra
freedom, without suffering the convergence issues of the unconstrained
model. The constraint on $\beta$ allows the majority of the values we found
when there was no constraint at all, but cuts off an extended wing that
reaches values as high as $\beta = 1.5 \times 10^{-12}\,$days.

Fig.~\ref{fig:p1p2_eq} shows the change in the $P_1$--$P_2$ MCMC projection as
the orbital models are given these greater freedoms. The changes are large,
showing that parameter degeneracy remains significant.  The orbital parameters
are consequently much more uncertain than the constrained model~2 of
\citet{Beuermann2010AA521L60} suggests, and it is no longer even clear whether
their model~2a (near 2:1) is favoured over 2b (5:2) as we see islands of
stability corresponding to both solutions. Perhaps most importantly however,
the increased model freedom allows access to long-lived parts of parameter
space, with significant regions of stability, somewhat allaying the
worry of the previous section over the likely complete disappearance of any
such models. This is particularly the case once the binary's period is
allowed to vary.

The means and standard deviations of the orbital parameters of models plotted
in Fig.~\ref{fig:p1p2_eq} are listed in Table~\ref{tab:params}, along with the
\begin{table*}
\caption{The first three columns list the means and standard deviations of the
  orbital parameters of the models shown in
  Fig.~\protect\ref{fig:p1p2_eq}. The model of
  the left-hand column uses the same fit parameters as
  \protect\citet{Beuermann2010AA521L60}'s model~2, hence the ``B'', with the
  next two adding the extra freedoms indicated. The right-hand column is the
  same as the left-hand one except only the pre-2011 data were used. The
  reference eclipse for the binary ephemeris, marked by $T_0$, is shifted
  forward by 43042 cycles relative to the usual ephemeris of NN~Ser to reduce
  the otherwise-strong correlation between $T_0$ and $P$.
  \label{tab:params}}
 
\begin{tabular}{lcccc}
\hline
Parameter & B   & B + $e_2$ & B + $e_2$ + $\beta$ & B \\
          & all & all & all & pre-2011\\
\hline
$T_0-52942.9338$ (MJD) & $(9.5 \pm 0.1) \times 10^{-5}$ & $(8.4 \pm 2.8) \times 10^{-5}$ & $(5.3 \pm 4.4) \times 10^{-5}$ & $(9.2 \pm 0.8) \times 10^{-5}$\\
$P-0.13008014$ (d) & $(2.4 \pm 0.1) \times 10^{-9}$ & $(2.3 \pm 0.3) \times 10^{-9}$ & $(2.7 \pm 0.5) \times 10^{-9}$ & $(1.8 \pm 2.6) \times 10^{-9}$\\
$\beta$ ($10^{-12}\,\mathrm{d}$) & --- & --- & $ 0.04 \pm 0.05$  & ---\\
$a_1$ (AU) & $ 3.488 \pm 0.012$  & $ 3.43 \pm 0.14$  & $ 3.37 \pm 0.15$  & $ 3.28 \pm 0.22$ \\
$P_1$ (yr) & $ 8.09 \pm 0.04$  & $ 7.9 \pm 0.5$  & $ 7.7 \pm 0.5$  & $ 7.4 \pm 0.8$ \\
$m_1$ ($\mj$) & $ 2.688 \pm 0.036$  & $ 2.3 \pm 0.5$  & $ 2.2 \pm 0.5$  & $ 2.2 \pm 0.9$ \\
$T_1$ (MJD) & $  58205 \pm 22$  & $  58106 \pm 228$  & $  58043 \pm 250$  & $  57826 \pm 378$ \\
$e_1$ & $ 0.163 \pm 0.007$  & $ 0.19 \pm 0.05$  & $ 0.19 \pm 0.04$  & $ 0.21 \pm 0.04$ \\
$\omega_1$ ($^\circ$) & $-107.4 \pm 2.7$  & $ -111 \pm 13$  & $ -118 \pm 15$  & $ -105 \pm 8$ \\
$a_2$ (AU) & $ 5.313 \pm 0.005$  & $ 5.35 \pm 0.06$  & $ 5.47 \pm 0.15$  & $ 5.51 \pm 0.18$ \\
$P_2$ (yr) & $ 15.125 \pm 0.021$  & $ 15.27 \pm 0.28$  & $ 15.8 \pm 0.7$  & $ 16.0 \pm 0.8$ \\
$m_2$ ($\mj$) & $ 7.46 \pm 0.05$  & $ 7.33 \pm 0.31$  & $ 7.29 \pm 0.32$  & $ 6.9 \pm 1.4$ \\
$T_2$ (MJD) & $ 53973.3 \pm 1.5$  & $  54016 \pm 106$  & $  54096 \pm 133$  & $  54008 \pm 58$ \\
$e_2$ & --- & $ 0.08 \pm 0.05$  & $ 0.09 \pm 0.05$  & ---\\
$\omega_2$ ($^\circ$) & --- & $  43 \pm 119$  & $  62 \pm 95$  & ---\\
$\chi^2$, $N_{dof}$ & 62.8, 66 & 62.6, 64 & 62.5, 63 & 31.8, 32 \\
\hline
\end{tabular}

\end{table*}
values corresponding to the lower-left panel of Fig.~\ref{fig:p1p2}. 
Most of the parameters have an obvious meaning, but it should be noted that
the epochs $T_1$ and $T_2$ refer to the time when the respective planet
reaches the ascending node of its orbit, not the more usual
periastron, as the latter is poorly defined for small eccentricities. The
eccentricity of the outer planet $e_2$ and the quadratic term in the binary
ephemeris $\beta$ are consistent with zero, although, as we have just seen,
dynamical stability seems to suggest that $e_2 > 0$, and it would not be
surprising were this the case. The $\chi^2$ values listed are the
minimum of any models of the MCMC chains. The MCMC method does not aspire to
find the absolute minimum $\chi^2$, and tests we have made suggest that the
values listed in the table are of order $0.5$ -- $1.5$ above the absolute
minimum. The improvement in $\chi^2$ as more parameters are added is marginal,
so a circular outer orbit is fine for fitting the data. It is the requirement
of dynamical stability which leads us to favour the model with eccentricity.
In using the numbers of Table~\ref{tab:params}, it should be realised
that the mean values do not need to correspond to any viable model: for
instance, the mean of a spherical shell distribution lies outside the
distribution itself.
 
The  quadratic term produced by a rate of angular momentum change $\dot{J}$ is
given by
\begin{equation}
 \beta = \frac{3 P^2}{2} \frac{\dot{J}}{J} ,
\end{equation}
where $P$ is the orbital period and $J$ is the angular momentum.  For the
parameters of NN~Ser \citep{Parsons2010MNRAS402.2591}, gravitational wave
radiation alone gives $\dot{J}/J = - 1.36 \times 10^{-18} \,\sec^{-1}$, and
therefore $\beta = - 3.0 \times 10^{-15}\,$days. Over the entire baseline of
observations of NN~Ser, the $\beta E^2$ term would then amount to $1.5\,\sec$.
Although in principle this is detectable, at present, because of the planets
(or whatever is causing the timing variability), there is strong degeneracy in
the fits once a quadratic term is allowed and we are far from being able to
measure a term this small. In fact, as we remarked earlier, the
degeneracy between $\beta$ and the outermost planet's orbital parameters is
so strong that $\beta$ is only weakly constrained by our data and the
uncertainty listed for $\beta$ in Table~\ref{tab:params} largely reflects
the prior restriction we placed upon it. The GWR prediction is the minimum
expected angular momentum loss, as one also expects some loss from magnetic
stellar wind braking. The secondary star in NN~Ser has a mass of
$0.111\,\msun$, making it comparable to short-period ($P \approx 90\,$mins)
cataclysmic variables for which there is evidence for angular momentum loss at
around $2.5 \times$ the GWR rate at the same short periods
\citep{Knigge2011ApJS194.28}, but this is still much smaller than we can
measure at present. We expect a substantial improvement in this constraint
over the next few years as the parameter degeneracy is lifted. Given
the current lack of constraint upon $\beta$ from the data, at present we
favour the model in which $\beta$ is fixed to zero.

\subsection{Comparison with 
\protect\citet{Beuermann2013AA555A.133B}}

\label{sec:new}

As mentioned earlier, shortly after the first submission of this paper,
\citet{Beuermann2013AA555A.133B} presented new eclipse times and a stability
analysis of NN~Ser. In this section we compare our sets of results which are
based upon the same set of data prior to 2011, but independent sets of new
data thereafter, i.e. we do not use any of their new
data. \citet{Beuermann2013AA555A.133B} consider only models equivalent to our
``$B + e_2$'' models of the middle panel of Fig.~\ref{fig:p1p2_eq}. They
fitted their data through Levenberg-Marquardt minimisation of $\chi^2$, which,
apart from the absence of prior probability factors, finds the region of
highest posterior probability, but does not explore the shape of region of
parameter space supported by the data as MCMC does. They imposed conditions of
dynamical stability, which makes a direct comparison with our results tricky
since we adopted the strategy of first seeing what parameter space was
supported by the data and only then testing dynamical stability. They found
stable orbits close to the 2:1 resonance if they allowed the orbit of the
outermost planet to be eccentric.  This is consistent with what we find: there
are almost no long-lived orbits if the outermost orbit is forced to be
circular, but some appear near the 2:1 line once eccentricity is allowed. We
refer to \citet{Beuermann2013AA555A.133B} for a detailed discussion of the
nature of the stable solutions that they find, in particular a demonstration
that they are in mean-motion resonance.  \citet{Beuermann2013AA555A.133B} did
not consider any period variation of the binary or explore the much wider
range of orbits this allows.  Thus they did not uncover any of the stable
models near the 5:2 ratio which are permitted by the data once period
variation is included, and therefore, although we agree that the 2:1 resonance
is favoured, we feel that their exclusion of the 5:2 resonance at ``99.3\%
confidence'' is premature.

\begin{figure}
\hspace*{\fill}
\includegraphics[width=\columnwidth,clip=]{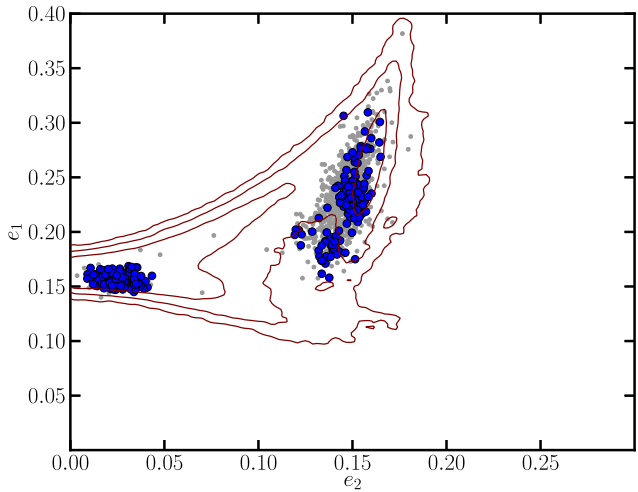}
\hspace*{\fill}

\caption{The projection onto the $e_2$--$e_1$ plane of the MCMC chain allowing
for eccentricity in both orbits but not binary period change, i.e. the models
shown in the central panel of Fig.~\protect\ref{fig:p1p2_eq}. The contours
show regions encompassing $68.3$, $95.4$ and $99.7$\% of the orbits  
supported by the data, with no restriction on stability. Small grey dots 
mark the orbits that last between $10^6$ and $50 \times 10^6$~years; large
blue dots mark those that last for more than $50 \times 10^6$~years. The
contours are comparable to the left panel of 
figure~3 from \protect\citet{Beuermann2013AA555A.133B}, while the locations of
the long-lived models are comparable to the other two panels of their figure.
\label{fig:e2e1_stab}}
\end{figure}

\citet{Beuermann2013AA555A.133B} present a plot of the dynamical lifetime as a
function of the eccentricities of the two planets, $e_1$ and $e_2$ (their
figure~3). This provides us with an opportunity to compare the constraints set
by our two sets of data, although as already remarked the differences between
our two approaches make exact comparison difficult. For instance, we reject
the implication of the right-hand two panels of their figure~3 that the
dynamical lifetime is a single-valued function of $e_1$ and $e_2$; instead,
once one allows for the distribution of other parameters, there must be a
distribution of lifetimes at any given values of $e_1$ and $e_2$; we discuss a
similar issue at length in the next section. However, a comparison can still
be made accepting that \citet{Beuermann2013AA555A.133B}'s figure shows the
lifetime of the most probable orbits, since for each $e_1$--$e_2$ point they
re-optimised the other 10 parameters. Our nearest equivalent to their plot is
shown in Fig.~\ref{fig:e2e1_stab}
for which we extended our dynamical
integrations to 100 million years to delineate regions of greatest long-term stability.
The figure compares well with figure~3 of \citet{Beuermann2013AA555A.133B}
with many similar features. We see the same tight definition of $e_1$ at low 
values of $e_2$, but spreading out as $e_2$ increases. The main island of
stable models found by \citet{Beuermann2013AA555A.133B} coincides with the
island of stable orbits that have high $e_2$ values seen in
Fig.~\ref{fig:e2e1_stab}.

There are a few differences as well. Our data support a smaller region of
parameter space, owing to a higher overall precision which more than
compensates for a smaller number of eclipse time measurements. In particular,
a spur of large $e_2$ / low $e_1$ allowed by
\citet{Beuermann2013AA555A.133B}'s data is eliminated by ours, and there is
general exclusion of high $e_2$ values leading to the large area of white
space on the right-hand side of the plot for which we chose the same axis
limits as \citet{Beuermann2013AA555A.133B}. We ascribe these differences to
signal-to-noise rather than anything more fundamental. The other most notable
difference is that we find an island of stability for $e_2 = 0.01$ -- $0.04$
as well. Although there are signs of the same region in
\citet{Beuermann2013AA555A.133B}'s figure, it is not as marked as we find.
This may be the result of the difference in approaches, with
\citet{Beuermann2013AA555A.133B} tracing the highest probability region for
each $e_1$--$e_2$ value, versus our exploration of the larger region of
supported parameter space.

These differences are small, and overall we conclude that we are in
substantial agreement with \citet{Beuermann2013AA555A.133B}. This is of course
to be hoped for given that we use the same data, with two small corrections,
up to 2011.

\section{Discussion}

The two-planet model for the variations in eclipse times of NN~Ser has
survived both new precise data and an updated dynamical stability analysis. It
is the first compact eclipsing binary apparently hosting planets for which
this can be said. It also delivers by far the highest quality eclipse times
with a weighted RMS scatter around the best fit orbit of $\sigma = 0.07\,$sec,
where
\begin{equation}
\sigma^2 = \frac{\chi^2 / (N-V)}{\left(\sum_{i=1}^N 1/\sigma_i^2\right) / N},
\end{equation}
with $N$ the number of data, $V$ the number of variable parameters, and
$\sigma_i$ the individual uncertainties on the eclipse times. The nearest
rival in this respect as far as we can determine is HU~Aqr for which
\citet{Gozdziewski2012MNRAS425.930} quote a scatter of $0.7\,$sec, and this
after significant pruning of discrepant points. Our typical best-fit values of
$\chi^2$ are around 63 with 76 points and 10 -- 13 fit
parameters. The expected value of $\chi^2$ is thus 63 to 66 $\pm 11$, so there
are as yet no signs of systematics in the data.

We have shown that the range of orbits consistent with
\citet{Beuermann2010AA521L60}'s data leads to a good prediction for the
location in the $O-C$ diagram of the new data, so the planet model has
predictive power. Moreover, allowing a non-zero eccentricity of the outer
planet's orbit, we find stable solutions. The latter result is interesting,
and perhaps counter-intuitive at first sight. One might expect if the outer
planet's orbit is allowed to be eccentric then it is more likely to
de-stabilise the orbit of the lighter inner planet. This is what
\citet{Horner2012MNRAS425.749} found, but we believe their analysis to suffer
from significant technical flaws. Some of these are common to other papers
from the same authors, as we now discuss.

\subsection{Previous dynamical stability analyses of NN~Ser and related systems}

\citet{Beuermann2010AA521L60} carried out a limited stability analysis of
NN~Ser's putative planetary system using $100$,$000\,\yr$-long integrations
and identified stable regions of parameter space, which they tentatively
associated with 2:1 and 5:2 mean-motion resonances.
\citet{Horner2012MNRAS425.749} pointed out that $10^5\,\yr$ was too short to
assess long-term stability, and also criticised the restriction to circular
orbits for the outer planet. They too found significant stability
when the outer planet was held in a circular orbit, but when they allowed its
eccentricity to vary and re-fitted the orbits, they found that the solution
lay within a broad region of very short-lived orbits, although uncertainties
were sufficient to allow for some long lasting ones too. They concluded this
from an examination of the lifetime of the system as a function of the
inner-planet's semi-major axis $a_1$ and eccentricity $e_1$
(their figure~5), and ascribed it to the significant eccentricity
($e_2 = 0.22$) they found for the outer planet's orbit. Our results
do not agree with theirs, and this is not simply to do with the new data,
because we still find significant numbers of stable solutions when we restrict
our analysis to the pre-2011 data used by \citet{Beuermann2010AA521L60} and
\citet{Horner2012MNRAS425.749}.

Instead, we believe that the work presented in \citet{Horner2012MNRAS425.749}
suffers from a series of flaws, the last of which renders it largely
irrelevant to the question of stability of NN~Ser. The same problem affects
a series of similar papers from the same authors, and thus we devote this
section to where we think this work has gone awry.

We start with minor issues.  First of all, NN~Ser is not, and never has been,
a cataclysmic variable, and, since its white dwarf is hot ($T_{eff} \approx
60$,$000\,$K, \citet{Wood1991ApJ381.551}), it only emerged from its common
envelope around one million years ago. This renders most of
\citeauthor{Horner2012MNRAS425.749}'s 100 million year-long integrations
superfluous since the system was undoubtedly very different prior to the
common envelope in a way that cannot be modelled with the Newtonian
dynamics of a few, constant point masses. Still, this does not alter
\citeauthor{Horner2012MNRAS425.749}'s claim of instability since they place
NN~Ser within a zone where orbits typically survive only $\sim 3000$
years. Another minor issue is that they used a total mass for NN~Ser of
$0.69\,\msun$ from \citet{Haefner2004AA428.181} rather than the more recent
determination of $0.646\,\msun$ from \citet{Parsons2010MNRAS402.2591} which
was used by \citet{Beuermann2010AA521L60}, thus they were not self-consistent
since they started from \citet{Beuermann2010AA521L60}'s solutions. Once more
however, this probably does not affect their essential claims. Their use of
astrocentric Keplerian fits, both from \citet{Beuermann2010AA521L60} and of
their own devising, are further drawbacks, because, as discussed earlier, no
Keplerian model is accurate enough to match the precision of the NN~Ser times,
and astrocentric coordinates perform worst of the three coordinate
parameterisations we examined.  However, our calculations indicate that this
should not have made a qualitative difference to 
\citeauthor{Horner2012MNRAS425.749}'s work either.

\begin{figure}
\hspace*{\fill}
\includegraphics[width=\columnwidth,clip=]{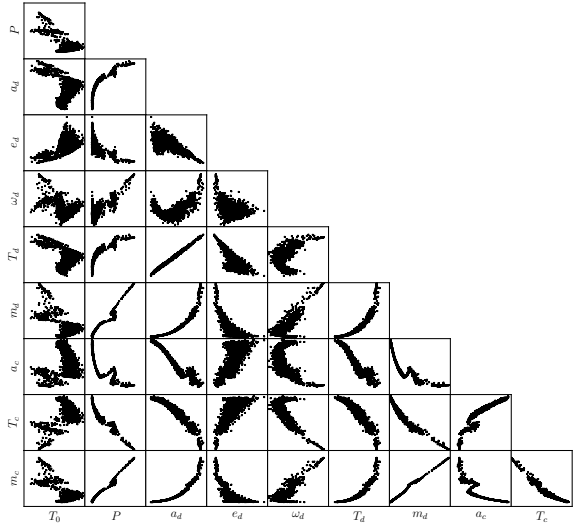}
\hspace*{\fill}

\caption{Scatter plots of the 10 parameters used during the MCMC runs shown
in the lower-left panel of Fig.~\protect\ref{fig:p1p2}. These are based upon
the pre-2011 data alone, with the outer planet held in a circular orbit and
$N$-body effects accounted for.
\label{fig:correlations}}
\end{figure}

This brings us to what we believe \emph{is} the major problem with
\citet{Horner2012MNRAS425.749}'s analysis, a problem which applies equally to
the series of papers from the same group analysing stability in related
systems. The figures upon which \citet{Horner2012MNRAS425.749} base their
conclusions show cuts through parameter space in which dynamical lifetime is
plotted as a function of two orbital parameters perturbed by $\pm 3 \sigma$ in
a grid around their best-fit values, variously the semi-major axis,
eccentricity and argument of periastron of the inner planet. The problem with
all of them is that they \emph{do not represent orbits consistent with the
  data} because in each case the remaining 10 free parameters have not been
adjusted. Correlations between orbital parameters are \emph{highly}
significant. Rather than slices through parameter space which very rapidly
fall out of the region supported by the data, what should be plotted are the
lifetimes of the \emph{projection} of models consistent with the data. In
general, as we indicated earlier when discussing
  Fig.~\ref{fig:e2e1_stab}, the result is not even a single-valued function of position in a 2D
projection, and it is quite possible to have very short- and very long-lived
models right on top of each other, an impossibility in
\citet{Horner2012MNRAS425.749}'s presentation. The MCMC method delivers just
what is needed through its generation of models which follow the posterior
probability distribution implied by the data. Fig.~\ref{fig:correlations}
displays all possible two-parameter projections of our MCMC models of the
pre-2011 data and shows complex and high-degree correlations between all
parameters. If anything, this figure undersells the problem since projections
from high- to low-dimensionality smear out correlations (imagine projecting a
spherical shell distribution from 3D to 2D for instance).  Failing to account
for these correlations is a serious error of methodology, and we believe it is
this which explains the difference between our results and those of
\citet{Horner2012MNRAS425.749}; Fig.~\protect\ref{fig:correlations} also makes
it clear that covariance matrix uncertainties based upon a quadratic
approximation to the minimum $\chi^2$ can under some circumstances be
extremely mis-leading.

\begin{figure}
\hspace*{\fill}
\includegraphics[width=0.8\columnwidth,clip=]{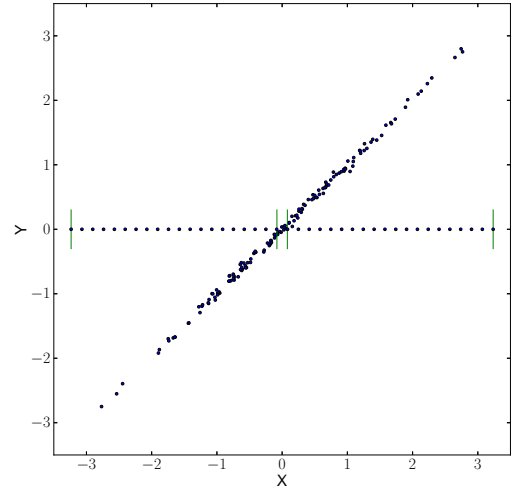}
\hspace*{\fill}

\caption{A schematic illustration of a serious problem with
  \protect\cite{Horner2012MNRAS425.749}'s stability analyses. The outermost
  vertical lines mark the $\pm 3 \sigma$ range in $X$ of the correlated set
  of points. This range is much larger than the range covering the
  intersection of these points with the $X$-axis, as indicated by the innermost
  pair of vertical lines. The regularly-spaced points along the $X$-axis which
  span the $\pm 3 \sigma$ range largely fall outside the region of the
  correlated points. The equivalents in
  \protect\cite{Horner2012MNRAS425.749} are the 2D grids over which
  they compute dynamical lifetimes; in the main these grids represent orbits
  which are incompatible with the data.
\label{fig:illustrate}}
\end{figure}

Fig.~\ref{fig:illustrate} 
presents a schematic illustration of the problem
with \citet{Horner2012MNRAS425.749}'s approach. It compares $\pm 3 \sigma$
range in $X$ of a set of points correlated in $X$ and $Y$ with the much
smaller zone where these points intersect the $X$ axis. Under this analogy,
\citet{Horner2012MNRAS425.749}'s method is the equivalent of choosing a set of
models that run along the $X$-axis over the $\pm 3\sigma$ range, as we show
with the regularly-spaced points in Fig.~\ref{fig:illustrate}. These
barely sample the region of the correlated
points; the problem can be expected to worsen with more dimensions. To assess
the scale of the problem in the specific case of NN~Ser, we calculated the
size of the 2D intersection in a plot analogous to
\citet{Horner2012MNRAS425.749}'s figure~5 which covers $\pm 3\sigma$ ranges on
the inner planet's semi-major axis and eccentricity, $a_1$ and $e_1$. When
just these two parameters are perturbed, we find that the $\chi^2$ minimum is
nearly quadratic. We thus defined the intersection as the region for
which $\chi^2 - \chi^2_\mathrm{min} < 13.8$ ($99.9$\% two parameter,
joint-confidence).  We found that the interaction cross-section occupies just
1 part in $10^4$ of the total area plotted. In other words $99.99$\% of the
area plotted by \citet{Horner2012MNRAS425.749} in their figure~5 is outside
the region of 10-dimensional parameter space supported by the data, just as
the regularly-spaced points in Fig.~\ref{fig:illustrate} are by-and-large 
outside the 2D distribution of points.

The problem with \cite{Horner2012MNRAS425.749}'s analysis of NN~Ser is of wide
impact since a very similar approach was applied to HU~Aqr by
\citet{Horner2011MNRAS416L11} and \citet{Wittenmyer2012MNRAS419.3258},
NSVS~14256825 by \citet{Wittenmyer2013MNRAS431.2150}, HW~Vir by
\citet{Horner2012MNRAS427.2812} and, most recently, to QS~Vir by
\citet{Horner2013arXiv1307.7893H}. In some cases these authors have averaged
the results over other parameters such as the mean anomaly and argument of
periastron of the particular planet orbit they perturb, but, as far as we can
determine, in no case do they allow for simultaneous variations of all other
fit parameters as is essential (and simply averaging over other parameters
fails to account for the weighting required to reflect the constraints of the
data in any case). We conclude that the issue of stability or instability in
these systems needs re-opening. It may well turn out that the conclusions of
this series of papers, which have for the most part found that proposed
multi-planet orbits around binaries are not dynamically viable, will remain
unchanged (we think it highly likely that the orbits proposed for QS~Vir are
unstable for instance), but some work is now required to be sure that
  this is the case.  This problem does not apply to the recent study of
NN~Ser by \citet{Beuermann2013AA555A.133B} because although their lifetime
versus $e_1$--$e_2$ plots are superficially similar to
\cite{Horner2012MNRAS425.749}'s plots, \citeauthor{Beuermann2013AA555A.133B}'s
optimisation of the other parameters ensures that they stayed in regions of
parameter space supported by the data.

\subsection{The immediate future of NN~Ser}
\begin{figure}
\hspace*{\fill}
\includegraphics[angle=270,width=\columnwidth,clip=]{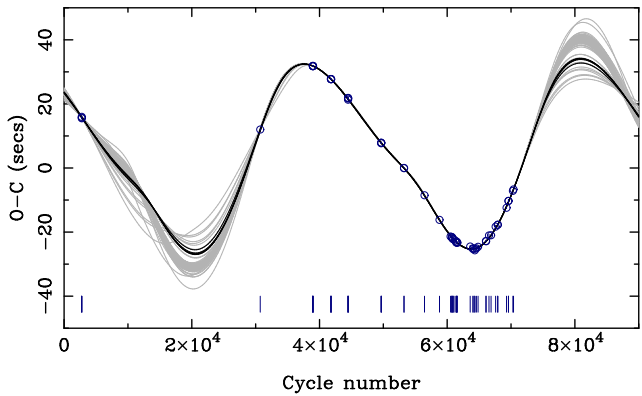}
\hspace*{\fill}

\caption{50 orbit fits to NN~Ser allowing for eccentricity in the outer
  planet's orbit diverge in the near future. Darker lines highlight
  those models which are stable for more than a million years. The reference
  ephemeris for this plot is $\mathrm{BMJD(TDB)} = 47344.0243673 +
  0.130080141716 E$. The plot extends until July 2020. Vertical lines at the
  bottom of the plot show the time sampling. Only points with uncertainties $<
  0.5\,$sec are shown.
\label{fig:future}}
\end{figure}

Since we have shown that the expected period change of the binary is much less
than our current measurement uncertainty, our favoured model for NN~Ser is one
in which we allow the outer planet's orbit to be eccentric, but do not allow
for any change in binary period, i.e. the middle set from
Fig.~\ref{fig:p1p2_eq}.  Using this set of models, Fig.~\ref{fig:future} 
shows all of the eclipse times of NN~Ser with uncertainties less than 0.5
seconds, and projects a few years into the future. We are still paying the
price for the poor coverage of the 1990s, but the next few years should see a
great tightening of the constraints. It appears from this plot that a sampling
interval of order a year or two should suffice.

\subsection{The planet hypothesis of eclipse timing variations}

Rather to our surprise, the new eclipse times of NN~Ser presented in this
paper are in good agreement with predictions based upon
\citet{Beuermann2010AA521L60}'s model in which two planets orbiting the binary
cause the timing variations. We say to our surprise, because if all eclipse
timing variations of compact binary stars are caused by planets, circum-binary
planets must be common, since when looked at in detail the majority show
timing variations \citep{Zorotovic2013AA549.95}. We have long worried, and
continue to worry, that the planet models are a glorified form of Fourier
analysis, capable of fitting a large variety of smooth variations. We may
simply have been lucky so far with NN~Ser that the ``orbits'' returned have
been stable, so, although our results are in line with the planet model, we do
not regard the question as to the reality of the planets to be settled
yet. Currently the main obstacle to a definitive answer is the
still-considerable degeneracy in the orbit fits. Continued monitoring will
cure this. However, it is notable that this degeneracy survives even with our
mean timing precision of around $0.07\,$secs. Since one would need
$\sim 200$ eclipse times of $1\,$sec precision to match a single time of $0.07\,$sec
precision, we require not just extended coverage, but extended precision
coverage. The ultimate goal should be to remove this degeneracy and, beyond
this, detect $N$-body effects.

The planet model for NN~Ser also survives the test of dynamical stability
which has cut down so many other claims. Although we have challenged the
methodology of many of these tests, we suspect that the general implication of
implausibly unstable orbits found for many systems will prove to be
correct. This is not the case for NN~Ser yet, although it perhaps might be
when further data are acquired, because the addition of new data has
consistently made it harder to locate long-lived solutions. Around 50\% of
viable orbits fitted to the data of \citet{Beuermann2010AA521L60} (with
circular outer orbits) were long-lived. With our new data, this dropped to
$0.02$\%, prompting us to allow for eccentric outer orbits. Even allowing for
eccentricity, we found a similar drop from $7.6$\% to $0.7$\% when we added
the two ULTRACAM points from July 2013.

\section{Conclusions}
We have presented 25 new high precision eclipse times of the close
white dwarf binary, NN~Ser. The new times impressively follow the increasing
delay predicted according to the two planet model presented by
\citet{Beuermann2010AA521L60}. Moreover, some of the models supported by the
full set of data are dynamically stable. We found during our analysis that the
difference between Keplerian and properly integrated Newtonian models is
significant compared to the data uncertainties and must be accounted for
during fitting, not just in follow-up dynamical analysis.

The new data substantially reduce the degree of degeneracy in the planet model
fits, but much still remains, especially if the models are given complete
freedom with eccentricity in both orbits and orbital period change of the
inner binary allowed. Such freedom may even be necessary as with the new data,
very few of the orbits with the outer planet constrained to have a circular
orbit are stable. With eccentricity allowed for both orbits we find orbital
periods of $7.9 \pm 0.5\,$yr and $15.3 \pm 0.3\,$yr, and
masses of $2.3\pm 0.5\,\mj$ and $7.3\pm 0.3\,\mj$, with
stable orbits having close to 2:1 and 5:2 period ratios. At present, if a
quadratic term is allowed in the binary ephemeris, degeneracy between it and
the outermost planet's orbit precludes an astrophysically significant
measurement of the period change of the binary; this should improve
significantly over the next few years.

Finally, we have demonstrated that several existing dynamical stability
analyses of NN~Ser and related systems are based upon a flawed methodology and
require revision.

\section*{Acknowledgments}
Dimitri Veras, Danny Steeghs, Peter Wheatley and Boris G\"ansicke are
thanked for conversations on the topic of this paper. We thank the referee,
Roberto Silvotti, for helpful comments. TRM and EB were supported under a
grant from the UK's Science and Technology Facilities Council (STFC),
ST/F002599/1. SPL and VSD were also supported by the STFC. SGP acknowledges
support from the Joint Committee ESO-Government of Chile. MRS is supported by
the Millenium Science Initiative, Chilean Ministry of Economy, Nucleus
P10-022-F. CC acknowledges the support from ALMA-CONICYT Fund through grant
31100025.

\bibliography{marsh}{}
\bibliographystyle{mn_new}

\label{lastpage}

\end{document}